\documentclass[conference]{IEEEtran}
\IEEEoverridecommandlockouts
\usepackage{cite}
\usepackage{amsmath,amssymb,amsfonts}
\usepackage{graphicx}
\usepackage{textcomp}
\usepackage{xcolor}

\renewcommand{\figurename}{Figure} 
\renewcommand{\thefigure}{\arabic{figure}}
\def\BibTeX{{\rm B\kern-.05em{\sc i\kern-.025em b}\kern-.08em
		T\kern-.1667em\lower.7ex\hbox{E}\kern-.125emX}}

\begin{document}
	\title{Covariance Self-Attention Dual Path UNet for Rectal Tumor Segmentation\\
		{\footnotesize Haijun Gao\textsuperscript{1}, Bochuan Zheng\textsuperscript{1,*}, Dazhi Pan\textsuperscript{1,*}, Xiangyin Zeng\textsuperscript{2}}
		\thanks{
			\textsuperscript{*}corresponding author: zhengbc@vip.163.com, padzzj@163.com.\\
			\textsuperscript{1}School of Mathematics \& Information, China West Normal  University, Nanchong, Sichuan, China. email: ghjcwnu@163.com. \\
			\textsuperscript{2}School of Computer Science, China West Normal  University, Nanchong, Sichuan, China. 
		}
	
	}
	
	\maketitle
	
	\begin{abstract}
		Deep learning algorithms are preferable for rectal tumor segmentation. However, it is still a challenge task to accurately segment and identify the locations and sizes of rectal tumors by using deep learning methods. To increase the capability of extracting enough feature information for rectal tumor segmentation, we propose a Covariance Self-Attention Dual Path UNet (CSA-DPUNet). The proposed network mainly includes two improvements on UNet: 1) modify UNet that has only one path structure to consist of two contracting path and two expansive paths (nam new network as DPUNet), which can help extract more feature information from CT images; 2) employ the criss-cross self-attention module into DPUNet, meanwhile, replace the original calculation method of correlation operation with covariance operation, which can further enhances the characterization ability of DPUNet and improves the segmentation accuracy of rectal tumors. Experiments illustrate that compared with the current state-of-the-art results, CSA-DPUNet brings 15.31\%, 7.2\%, 11.8\%, and 9.5\% improvement in \textit{Dice} coefficient, \textit{P}, \textit{R}, \textit{F1}, respectively, which demonstrates that our proposed CSA-DPUNet is effective for rectal tumor segmentation.
	\end{abstract}
	
	\begin{IEEEkeywords}
		Covariance Self-Attention, UNet, Rectal Tumor Segmentation, convolutional neural network
	\end{IEEEkeywords}
	
	\section{Introduction}
	Rectal cancer is one of the most common malignant tumors of the digestive tract \cite{b01}, which is commonly formed between the dentate line and the junction of the sigmoid colon of the rectum. According to the 2020 cancer statistics of US, rectal cancer ranks the second among the most threatening cancer with a large number of patients and a high mortality rate \cite{b02}. Fortunately, there is a relatively slow pathological process from rectal tumors to rectal cancer, then it can be effectively prevented as long as we find and remove it in time \cite{b03}. Therefore, the precise identification and positioning of early rectal tumors is of great significance to the diagnosis and treatment of rectal cancer. In general, CT (Computed Tomography) image is a key technology for diagnosing the tumor or potential risk of carcinogenesis in the rectum. However, it is because of technical limitations and the indistinct characteristics of rectal tumors that rectal tumors are not easy to be observed in the CT images. If the doctors are not experienced and careful enough, the rectal tumors are easy to be unnoticed and misdiagnosed. With the rapid development of artificial intelligence algorithm-based technology, medical image segmentation method derived from image processing has become a research hotspot. As a complex and critical step in the field of medical image processing and analysis, medical image segmentation aims at segmenting medical images with certain special meanings and extract relevant features, providing a reliable basis for clinical diagnosis and pathology research, assisting doctors to make more accurate diagnosis, it has achieved great results in medical diagnosis.\par
	
	Medical image segmentation is an important part of artificial intelligence assisted medical technology, it has become a
	key research task to help doctors identify and diagnose major diseases. Medical image segmentation derives from semantic segmentation, and many effective algorithms have been proposed by researchers in recent years, such as Fully Convolutional Networks (FCN) \cite{b04}, Mask RCNN \cite{b05}, UNet \cite{b06}, UNet++ \cite{b07}, DeepLab v1-3 [8-10], nnUNet \cite{b11}, UNet 3+\cite{b12}, etc. The methods [3-11] are image segmentation methods based on convolutional
	neural network (CNN), and have achieved state-of-the-art results. UNet \cite{b06} is a classic work for medical image segmentation. It consists of an encoder network (for semantic information extraction) and a decoder network, and the nodes at same level between the encoder and the decoder are inter-connected by a jump connection to fuse the deep and shallow semantic information for enhancing the characterization of the medical image segmentation model. UNet++ \cite{b07} re-design skip pathways that connect the encoder and decoder networks and adopt deep supervision on the basis of UNet to further improve the segmentation accuracy of the model. nnUNet \cite{b11} proposed a robust adaptive framework based on 2D UNet and 3D UNet, which is no need to adjust parameters manually. Moreover, nnUNet does not use residual ResNet \cite{b13}, Dense connection, attention mechanism [14, 15, 16, 17] related skills, but make improvements in the methods of data preprocessing, training, inference strategies and post-processing. Although these works have shown state-of-the-art performance, they are still unable to obtain enough good accuracy for rectal tumor segmentation. So, we plan to modify the network model to explore more effective semantic information from a comprehensive range. Inspired by UNet++ and nnUNet, we proposed a new UNet network model, different from one contracting path and one expansive path of UNet++ and nnUNet, the proposed model consists of two contracting paths and two expansive paths. We name the new UNet network model as DPUNet(meaning Dual Path UNet). DPUNet not only integrates multi-scale semantic information, but the decoder can obtain sufficient fine-grained and coarse-grained semantic information at multiple scales. Besides improving detecting accuracy, it can also reduce network parameters and improve computational efficiency.
	\par
	
	Global modules based on long-range information dependency between pixels are widely used in semantic segmentation tasks, including the medical image segmentation. To better understand this dependency relationship, we use a self-attention mechanism \cite{b18} to mathematically model the dependency relationship between pixels. In some detail, the self-attention mechanism is aimed at characterizing a second-order relationship using a dot product operation for any two pixels features. However, the correlation between two pixels cannot be completely reflected by the result of dot product operation. Since each pixel contains rich information, besides its own position and pixel value information, the influence it has made on other pixels in the image should also be considered. And the conventional dot product operation usually only considers the influence of the pixel itself and ignores the influence of the global information, so its characterization is not rigorous and comprehensive enough. Considering this background, we calculate the covariance between pixels to replace the conventional dot product operation when calculate the correlation of the self-attention mechanism. On the other hand, the original self-attention mechanism requires big space to store attention map, so we adopt the criss-cross attention network proposed in CCNet \cite{b19} to solve this problem. The criss-cross attention network constructs a cross operation mode to calculate the correlation between pixels, which helps to reduce the space complexity of the attention feature map. Since our self-attention model is based on covariance, we call our rectal tumor segmentation network as CSA-DPUNet (meaning Covariance Self-Attention Dual Path UNet). 
	\par
	
	The remainder of this paper is organized as follows. In Section II we briefly review related methods for medical image segmentation and attention mechanism. Our proposed methods are given in Section III, where we describe two improvements which can incease the performance of rectal tumor segmentation. The experimental results are shown in Section IV. Finally, Section V summarizes our conclusions.
	\section{Related Works}
	
	\subsection{Medical Image Segmentation}
	According to the relevant literatures, a variety of segmentation models have been greatly developed based on CNN, such as fully convolutional neural networks (FCN) \cite{b04}, UNet \cite{b06}, PSPNet \cite{b20} and a series of DeepLab v1-v3 [8-10]. Especially, UNet established on the architecture of encoder-decoder is widely used in medical image segmentation. UNet uses skip connections to combine the image high-level semantic feature of the decoder and the image low-level detailed feature of the encoder. Aiming to reduce the semantic gap between the encoder and decoder, UNet++ \cite{b07} further enhanced these connections by introducing nested and dense skip connections. In terms of different information could be explored in different proportions of feature maps, a series of extended models based on UNet, including YNet \cite{b21}, UNet++ \cite{b07}, Attention UNet \cite{b22}, FocusNetAlpha \cite{b23}, UNet3+ \cite{b12}, etc. have been proposed. In some detail, the low-level detailed feature map can capture a wealth of spatial information, so that the boundaries of the organ can be highlighted; while the high-level semantic feature map reflects the position information, in which the location of the organ could be observed. Referring to the U-shaped architecture, Huang et al. \cite{b12} proposed a novel model UNet 3+ that reconstructs the connections between the encoder and the decoder and internally. The role of the connections between decoders is to capture fine-grained details and coarse-grained semantics from the entire scale. At the same time, in order to further learn the representation of the hierarchical structure from the full-scale aggregate feature map, the output of each side is connected to the mixed loss function, which helps for accurate segmentation, especially for organs that appear with different proportions in the medical images. However, as the sampling rate is reduced or increased, these precise segmentation signals may be gradually diluted.
	\subsection{Attention Mechanism}
	As a computing resource allocation scheme, attention mechanism uses limited resource allocation to process more important information to solve the problem of information overload. Generally, the input of neural network often contains a lot of redundant information, such as images and voice, not all information needs to be focused on, so we can play attention only to something important to improve time and space utilization.
	\par
	Therefore, there is an endless stream of works are carried out to incorporate attention mechanisms into deep neural networks to improve model learning capabilities. It has gained great application in the field of natural language processing (NLP), computer vision and recommendation systems. In computer vision, attention mechanism is used to teach the network learning to select some important region information for the next training while that irrelevant information is ignored. Bengio et al. \cite{b24} proposed a long-range information interaction model based on semantic information, the influence of the attention mechanism can be concentrated on the important areas of contextual semantic information. Vaswai et al. \cite{b14} proposed a network structure (Transformer) that is composed of a series of attention mechanisms, including spatial attention, channel attention, pixel attention, multi-head attention and self-attention, etc. [15, 16, 17]. The Squeeze-and-Excitation Networks (SENet) model \cite{b25} adopt the channel attention mechanism. Compared with the spatial attention mechanism, the channel attention mechanism pays more attention to the allocation of resources amonge convolution channels. Each convolution kernel in the CNN has corresponding feature channel, and the characterization capability of network is enhanced by calculating  between channels. In \cite{b17}, the dual-channel model DANet combines with spatial attention and channel attention. By the means of adding attention mechanism to the space and channel of the network, DANet performs well in semantic segmentation.
	\par
	The self-attention mechanism is also called the internal attention mechanism, it is designed to calculate the corresponding semantic information through modeling the relationship between different positions in a single sequence, and then directly model the long-range information interaction between semantics. According to the position invariance of the image, the encodings of all semantic information are structurally the same, but the semantic information parameters are not shared. The spatial attention mechanism makes the network pay more attention to the spatial position of the target, and the channel attention mechanism tends to focus on the size of the target \cite{b17}. The self-attention mechanism just concerns single rather than multiple cross-modal semantic information, that is, query, key and value are all obtained from the same semantic information \cite{b15}. Wang et al. \cite{b26} proposed a soft attention mechanism based on the idea of residual network to increase the diversity and quality of attention feature maps. Through expanding the size of the convolution kernel and introducing global information, the non-local network can obtain long-range of the receptive field in CNN. Wang et al. \cite{b27} proposed a non-local module that generates an attention map by calculating the correlation matrix between each spatial point in the feature map, and then aggregates contextual semantic information through the attention map. Consequently, OCNet  \cite{b28} and DANet  \cite{b17} use the non-local module to collect contextual information. The idea of PSANet  \cite{b29} is to let the network learn the attention map by adaptively learning each point assembles contextual information. Chen et al.  \cite{b30} used several attention masks to fuse feature maps or prediction maps from different branches. Lin et al. \cite{b31} believes that the attention established based on the dot product model has the characteristics of coupling. Thus authors splits the dot product attention through a disentangled method, the dot product is divided into a whitened pairwise term and a unary term. The disentangled non-local networks introduces non-local network into attention mechanism, and achieved state-of-the-art performances in the semantic segmentation network.
	\par
	Because of the advantages of the attention mechanism, it has also been frequently used in biomedical images. Oktay et al. \cite{b22} proposed an Attention-UNet network for medical image segmentation. Kaul et al. \cite{b23} integrated the attention of the gating mechanism into the UNet network to obtain a biomedical image segmentation network--FocusNetAlpha. Guo et al. \cite{b32} proposed the SA-UNet network, which embed the spatial attention into the a UNet network model and have been successfully applied for retinal segmentation. Ding et al. \cite{b33} proposed the hierarchical attention network HANet for medical image segmentation, which divided the attention mechanism into two-layer structure for feature fusion, and state-of-the-art segmentation effect was presented in the LUNA and DRIVE datasets. The common point of the above algorithms is to increase the characterization ability of network and extract more abundant and effective information by integrating the attention mechanism into methods. We aim to embed the self-attention module into the segmentation network, and use a self-developed correlation operation between pixels to calculate the attention map, making the network learn the attention information better.
	
	\section{Methods}
	This section introduces the basic structure of DPUNet and the covariance self-attention block.
	\subsection{Daul Path UNet}
	The basic structure of DPUNet is shown in Figure 1, which mainly including 3 sub-modules: each yellow circle represents a basic Block X, which is a standard convolution block; the blue arrow represents the down-sampling operation; the red arrow represent the up-sampling operation, which adopt deconvolution operation. Obviously, the mainly difference from UNet, UNet++ and nnUNet is that the proposed DPUNet has two contracting paths and two expansive paths, which are expected to not only integrate multi-scale semantic information, but also obtain sufficient fine-grained and coarse-grained semantic information at multiple scales.\par
	\begin{figure}[htp]
		\centering
		\includegraphics[scale=0.35]{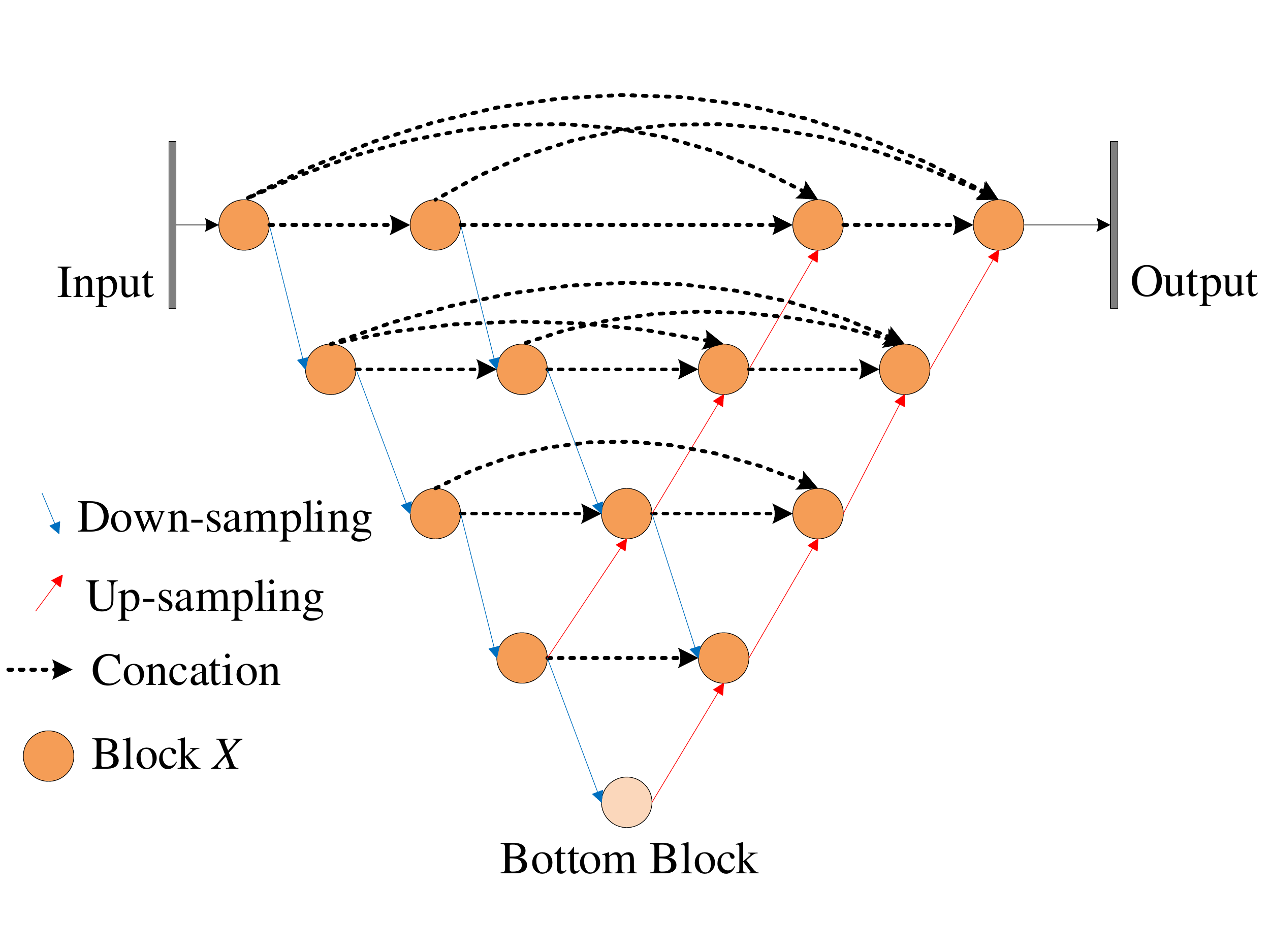}
		\caption{The illustration of DPUNet for rectal tumor segmentation. In this example, the input has 1 channel and the segmentation task has 1 class. In this model, two different operation modules are used, as shown in Figure 2 (a)-(c).}
		\label{fig1} 
	\end{figure}
	The details of the 3 sub-modules are listed in Figure 2, each sub-module has residual structure [13] for increasing the representational capability. The convolution block (shown in Figure 2(a)), which is adopted by all Block Xs and Bottom Block, has two convolution layers and two BN+ReLU layers, and obtains same size of feature map as input. The structure of the down-sampling module (shown in Figure 2(b)) is similar to the convolution block (shown in Figure 2(a)), their different is that one $3\times3$ convolution layer of the down-sampling module has a stride of 2. Therefore, the size of output feature map obtained by down-sampling module is only one half of the input feature map. To adjust the spatial sizes of feature maps accordingly, a $1\times1$ convolution with a stride of 2 is used to replace the identity residual connection. To remain enough feature information, the channel number of output feature map is twice as much as that of the input. The up-sampling sub-module of DPUNet (shown in Figure 2(c)) uses deconvolution operation to replace the identity residual connection for matching with the feature map size of up-sampling.\par
	
	\begin{figure}[htp]
		\centering
		\includegraphics[scale=0.25]{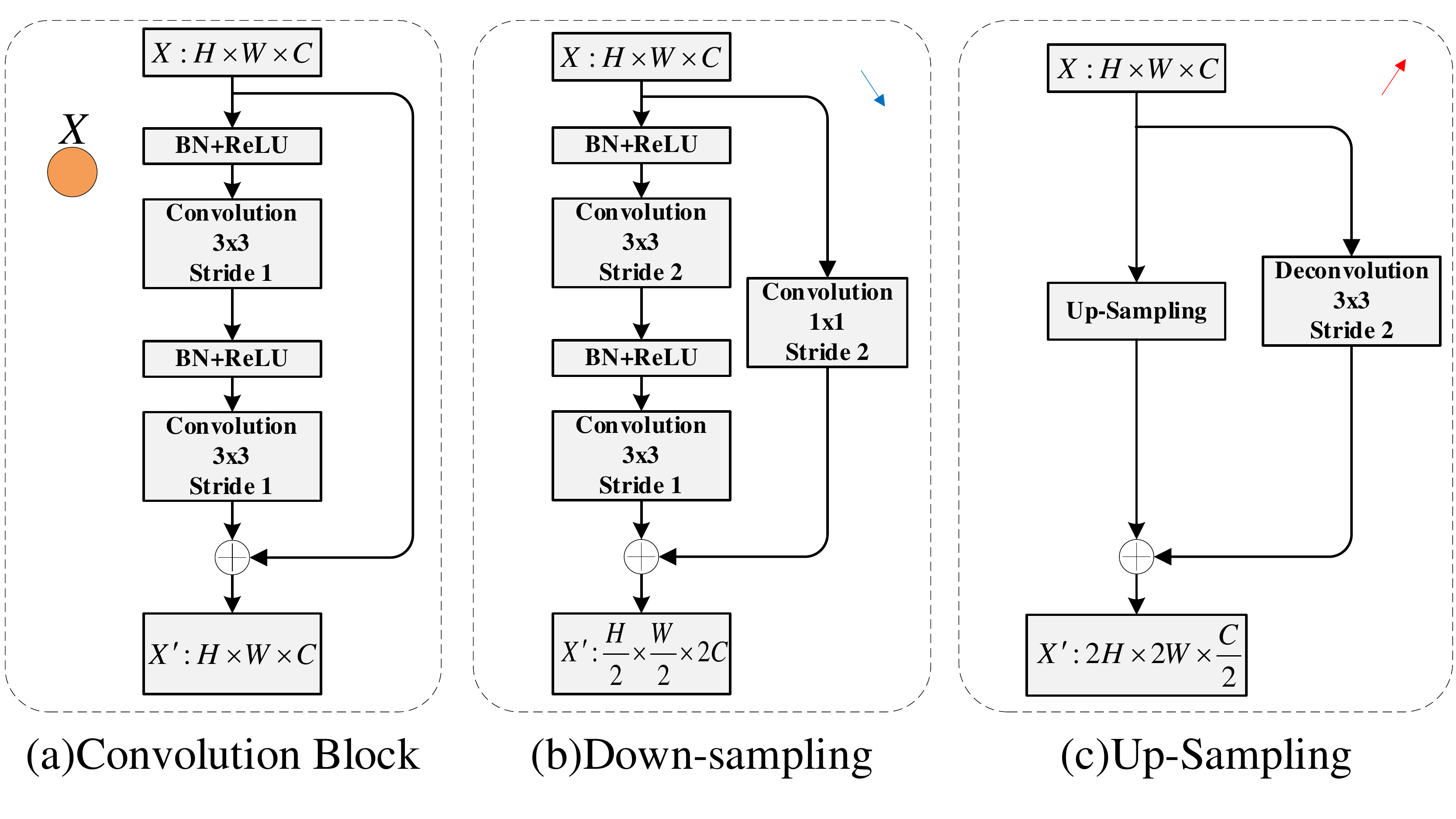}
		\caption{ The illustration of each modules employed by DPUNet.}
		\label{fig2} 
	\end{figure}
	
	To improve the performance of DPUNet, we further modify the structure of DPUNet, make it have the function of covariance self-attention. The covariance self-attention block is embedded into the up-sampling and bottom block sub-modules, which are illustrated in Figure 3. Consequently, we name the new DPUNet as CSA-DPUNet (meaning covariance self-attention DPUNet). For up-sampling sub-module, a covariance self-attention block is added into it after Up-Sampling block, shown as Figure 3(a). For bottom block, we simplify the original structure, as shown in Figure 3(b), that is a residual connection is applied on only the covariance self-attention block, so, the simplifed structure is different from the original structure (shown in Figure 2(b)).\par
	
	\begin{figure}[htp]
		\centering
		\includegraphics[scale=0.3]{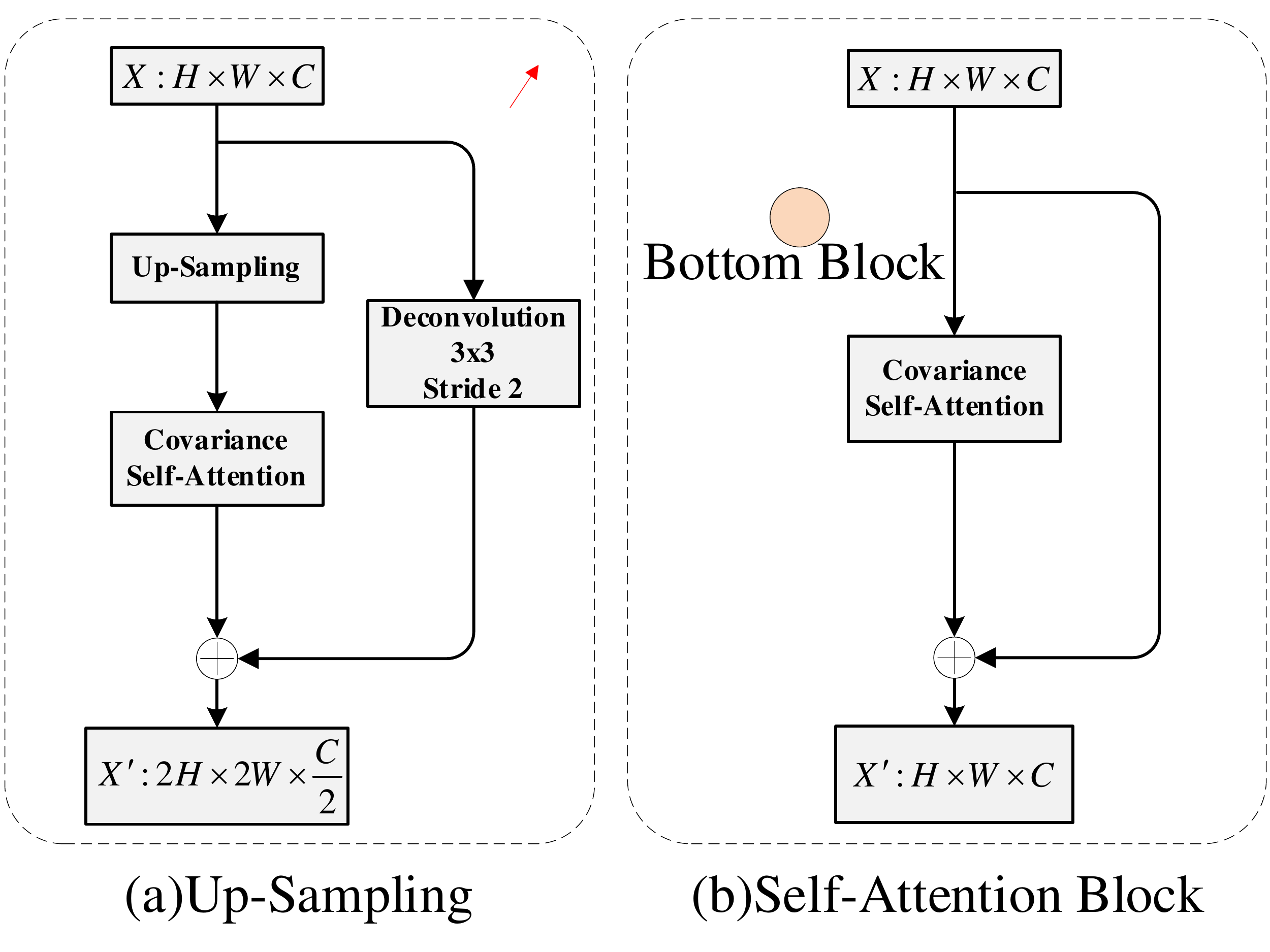}
		\caption{ The illustration of the modified modules.}
		\label{fig3} 
	\end{figure}
	
	\subsection{Covariance Self-Attention Block}
	This section introduces covariance self-attention blcok, as shown in Figure 4. Given a feature map $X\in \mathbb{R}^{h\times w\times d}$, the covariance attention block firstly use two convolution layers with $1\times 1$ filters to generate two feature maps $Q$ and $K$, respectively, where $Q\in \mathbb{R}^{h\times w\times d_q}$ and $K\in \mathbb{R}^{h\times w\times d_k}$,  $d_q$ and $d_k$ are the channel numbers of feature maps $Q$, $K$, respectively, and $d_q=d_k$ is less than $d$  for dimension reduction. The self-attention feature maps $SA\in \mathbb{R}^{h\times w\times (h+w-1)} $ are generated by feature maps $Q$ and $K$ through correlation operation. At each position $u$ in the spatial dimension of feature maps $Q$, we can get a vector $Q_u\in \mathbb{R}^{d_q}$. At each corresponding spatial position $u$ feature maps $K$, the correlation operation is defined as follows:
	\begin{equation}
	C_{i,u}=\left(Q_u-\bar{Q}_u \right)\left(K_{i,u}-\bar{K}_{i,u} \right)^T
	\end{equation}
	\par
	Where $K_{i,u}$ represents the $i$-th element of $K_u\in \mathbb{R}^{d_k\times (h+w-1)}$, $\bar{Q}_u$ is the mean of $Q_u$, $\bar{K}_{i,u}$ is the mean of $K_{i,u}$, and $i=[1, 2, \cdots, h+w-1]$. Obviously, $C_{i,u}\in \mathbb{R}$ is the covariance between feature vector $Q_u$ and
	$K_{i,u}$, thus convariance matrix $C\in \mathbb{R}^{h\times w \times (h+w-1)}$. After we obtain $C$ through eq.(1),
	we calculate self-attention map $SA$ from matrix $C$ by using $softmax$, $SA$ has same size as matrix $C$.
	\begin{equation}
	SA=softmax\left( C\right) 
	\end{equation}\par
	The feature map $V\in \mathbb{R}^{h\times w\times(h+w-1)}$ is obtained by convoluting the input $X$ with $1\times 1$ fiters, as shown in Figure 4. At each position $u$ in spatial dimension of feature map $V$, we can obtain a cross set 
	$\Phi_u\in \mathbb{R}^{d_v\times (h+w-1)}$ from $V$, $d_v=d$. Then, the long-range
	contextual information dependency fusion operation between pixels is defined as:
	\begin{equation}
	H_u=\sum_{i=1}^{h+w-1}SA_{i,u}\Phi_{i,u}
	\end{equation}\par
	Where $H_u$ is the feature vector of the output feature map $H\in \mathbb{R}^{h\times w\times d}$ at position $u$, 
	$SA_{i,u}$ is a scalar value at $i$-th channel and position $u$ of $SA$. In this way, the contextual semantic
	information is aggregated according to the spatial attention map.
	\begin{figure}[htp]
		\centering
		\includegraphics[scale=0.24]{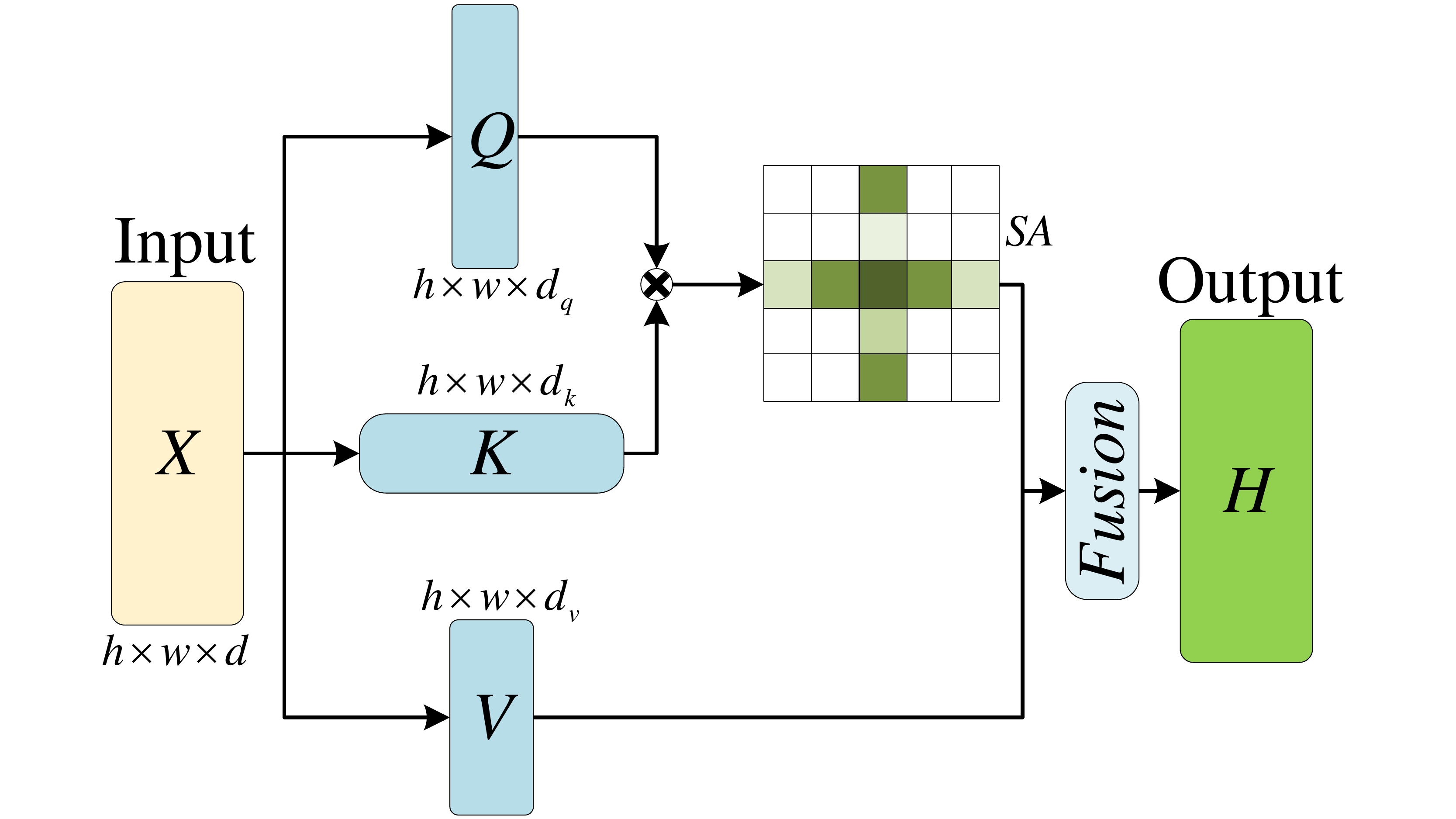}
		\caption{The details of criss-cross attention model. Each box is an input (X), output (H), intermediate matrix, respectively. the size of the corresponding matrix is outside each box. The variable name the of the matrix are inside each box. The black cross denotes correlation operation.}
		\label{fig5} 
	\end{figure}
	\par
	From above caculation of covariance self-attention feature maps, the main different from [19] is correlation operation. We adopt the method of caculating covariance matrix to caculate the correlation between pixels. Therefore, we call our self-attention block as covariance self-attention block.
	
	\par
	
	\section{EXPERIMENTS}
	In this section, we do a series of experiments to prove the effective of the proposed CSA-DPUNet. All experiments are conducted on the rectal tumor segmentation dataset obtained from [1].
	\subsection{Rectal Tumor Datasets}
	In this paper, we use the rectal tumor dataset provided by the China Seventh Teddy Cup Mining Challenge Competition to train and test our network and the other compared networks. The all images in dataset are CT images of rectum taken from CT scanning machine. The size of each image is $256\times 256$. The rectal tumor dataset contains 1693 images for training, each image has at least a rectal tumor area. The test dataset has 104 images, of which, each of 102 images contains at lease one rectal tumor area, while 2 images have no. The patients whose CT images contained in the test dataset have no CT images in the training dataset. No matter adopting which rectal tumor segmentation network to experiment, we all adopt same methods of pre-processing and data augmentation as in \cite{b34} to process the original data.
	\par
	Table 1 summarizes our results for the experiments. DPUNet and CSA-DPUNet outperforms every architecture across all metrics significantly for the dataset.
	
	\subsection{Evaluation Metric}
	In our work, the predicted segmentation masks are evaluated by the evaluation metrics. To quantitively evaluate the performance of rectal tumor segmentation network, four metrics, including Dice coefficient ($Dice$), Precision ($P$), Recall ($R$), F1 score ($F1$), are defined as the following:
	
	\begin{equation}
	Dice = \dfrac{2TP}{2TP+FP+FN}=\dfrac{2\left| A\cap B \right| }{\left| A\cup B \right|}
	\end{equation}
	
	\begin{equation}
	P=\dfrac{TP}{TP+FP},R=\dfrac{TP}{TP+FN},
	F1=\dfrac{2\times P\times R}{P + R}
	\end{equation}
	Where TN and TP denote the pixel number of true negatives and true positives, respectively; FN and FP denote the pixel number of false negatives and false positives, respectively; A and B are the predicted and ground truth of rectal tumor area, respectively. The higher of the values of  four metrics, the better the performance of rectal tumor segmentation network.
	\subsection{Results}
	The results of comparing experiments of 8  rectal tumor segmentation networks  are shown in Table 1. Obviously, the performance of DPUNet and CSA-DPUNet are better than the other compared networks on all four metrics, segmentation performance is improved significantly. Compared with the U-Net-SCB \cite{b34}, which is the best among the other compared networks, CSA-DPUNet brings 15.31\%, 7.2\%, 11.8\%, and 9.5\% improvement in \textit{Dice} coefficient, \textit{P}, \textit{R}, {\textit{F1}, respectively. Meanwhile, CSA-DPUNet obtains the best evaluation metric values except \textit{P}, which illustrates that the covariance self-attention module can improve the performance of DPUNet.
		\begin{table}[htbp]
			\caption{Comparing with  different  networks on rectal tumor dataset.}
			\begin{center}
				\centering
				\begin{tabular}{|c|c|c|c|c|}
					\hline
					\textbf{Methods} & \textbf{\textit{P}}& \textbf{\textit{R}}& \textbf{\textit{F1}} & \textbf{\textit{Dice}} \\
					\hline
					\hline
					UNet[6]& 71.63&69.24 &70.41&61.86  \\
					\hline
					Y-Net[21]& 78.21&76.81 &77.50&71.96  \\
					\hline
					UNet++[7]& 79.98&79.78 &77.50&79.61  \\
					\hline
					Attention UNet[22]& 76.26	&67.45	&76.85	&74.85  \\
					\hline
					FocusNetAlpha[23]& 81.22&79.36&80.28&81.60  \\
					\hline
					U-Net-SCB [34]	&89.84	&87.85&	88.83&	83.12\\
					\hline
					DPUNet	& \textbf{99.51}	&93.61	&96.47	&96.35\\
					\hline
					CSA-DPUNet	&97.04&	\textbf{99.65}	&\textbf{98.33}&\textbf{98.43}\\
					\hline
				\end{tabular}
				\label{tab1}
			\end{center}
		\end{table}
		\par
		\begin{figure*}
			\centering
			\includegraphics[scale=0.7]{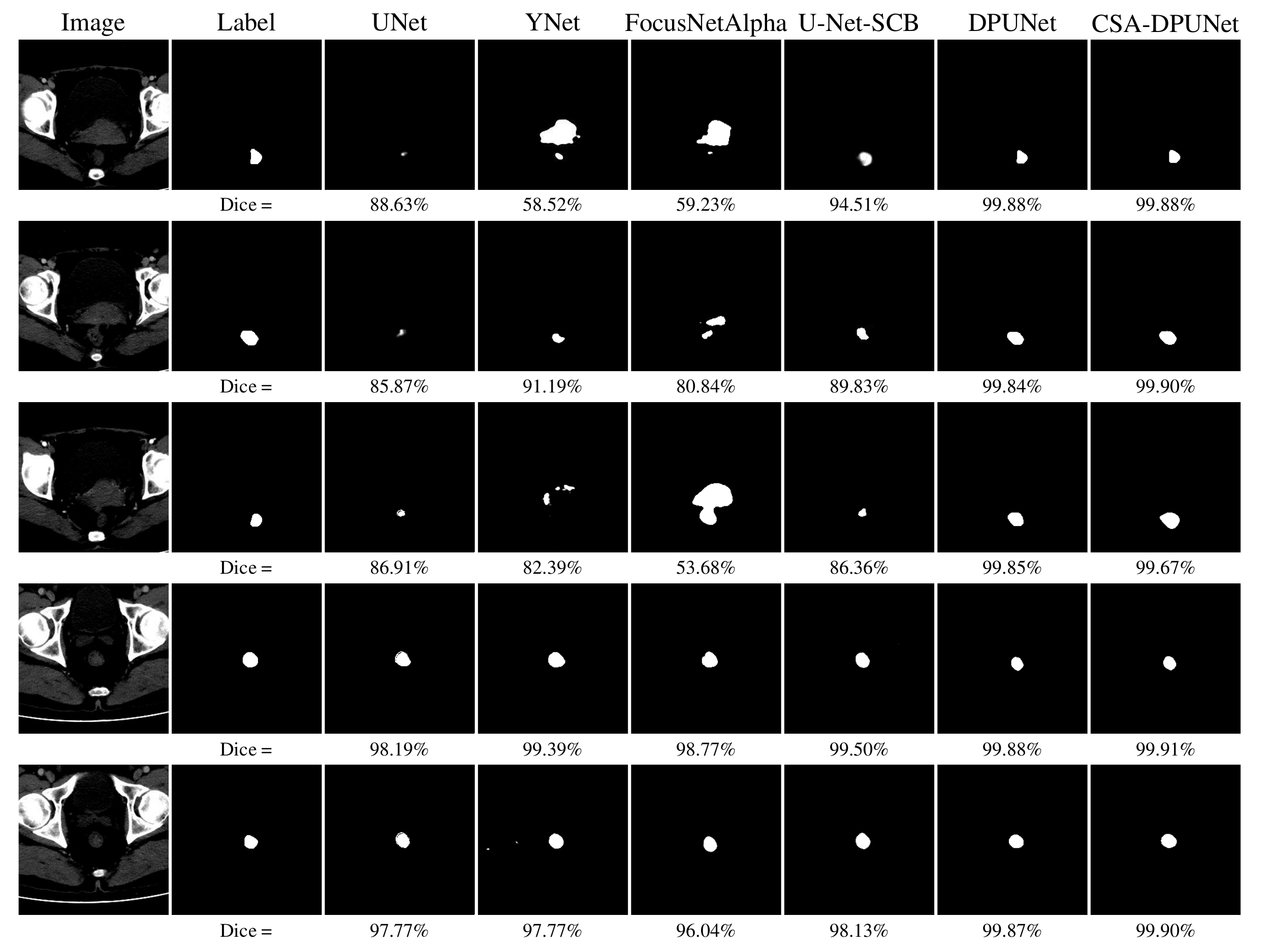}
			\caption{The comparing segmentation results of 5 CT images. The corresponding \textit{Dice} value is listed at the bottom of each segmentation images.}
			\label{fig55}
		\end{figure*}
		Figure 5 illustrates segmentation results of 5 CT images obtained by UNet \cite{b06}, YNet \cite{b21}, FocusNetAlpha \cite{b23}, U-Net-SCB \cite{b34}, DPUNet and CSA-DPUNet, respectively. Each row is corresponding  one CT image,  first column is original CT image, second column is ground truth, from column 3 to 8 are segmentation results. Obviously, the segmentation results of DPUNet and CSA-DPUNet is mostly similar to the ground truth, and the \textit{Dice} value listed in the bottom of each segmentation image is also support this conclusion. Although both DPUNet and CSA-DPUNet have achieved state-of-the-art performance, CSA-DPUNet is better than DPUNet, as shown in Figure 5.\par
		
		\begin{table}[htbp]
			\caption{Comparing with  different  attention on rectal tumor dataset.}
			\begin{center}
				\centering
				\begin{tabular}{|c|c|c|c|c|}
					\hline
					\textbf{Methods} & \textbf{\textit{P}}& \textbf{\textit{R}}& \textbf{\textit{F1}} & \textbf{\textit{Dice}} \\
					\hline		
					\hline
					CC-DPUNet&\textbf{97.63}	&94.68	&96.13	&96.06\\
					\hline
					SA-DPUNet&	97.04&	97.79&	97.71&	97.42\\
					\hline
					CSA-DPUNet	&97.04&	\textbf{99.65}	&\textbf{98.33}&\textbf{98.43}\\
					\hline
				\end{tabular}
				\label{tab2}
			\end{center}
		\end{table}
		In order to demonstrate the effective of the covariance self-attention block, we train and test both DPUNet embedded original self-attention mechanism \cite{b19}, the criss-cross attention \cite{b19} and CSA-DPUNet. We call DPUNet embedded the original self-attention block as SA-DPUNet and the criss-cross self-attention block as CCSA-DPUNet. SA-DPUNet and CCSA-DPUNet both use dot product operation to calculate correlation between two pixel vectors. The experimental results are shown in Table 2. As can be seen from Table 2, compared with CCSA-DPUNet, CSA-DPUNet is increased by 2.37\%, -0.59\%, 4.97\%, and 2.37\% in \textit{Dice} coefficient, \textit{P}, \textit{R}, {\textit{F1}, respectively. Compared with SA-DPUNet, CSA-DPUNet is increased by 1.01\%, 0\%, 1.86\%, and 0.62\% in \textit{Dice} coefficient, \textit{P}, \textit{R}, {\textit{F1}, respectively. So, the evaluation metrics of CSA-DPUNet are all better than that of SA-DPUNet and CCSA-DPUnet, which demonstrate the covariance calculation can improve the performance of SA-DPUNet.
						
						\section{Conclutions}
						In this paper, we propose a new rectal tumor segmentation network having two contracting paths and two expansive paths, as well as, improve the self-attention block and embed it into the network. The experimental results demonstrate that our  rectal tumor segmentation network can obtain state-of-the-art performance for the rectal tumor segmentation. Rectal cancer is one of the most serious malignant tumors in the world, with a high mortality rate. The accurate segmentation of the rectal tumor area is the key for the early diagnosis of rectal cancer. In this paper, the proposed deep learning method of rectal tumor segmentation has high precision, which can assist doctors in the diagnosis and recognition of rectal tumors, and helps reduce missed diagnosis and reduce the intensity of doctors' work. Auto segmenting rectal tumor area by using deep learning method is an important application of artificial intelligence technology in the field of medical imaging, and is also an important part of smart medical treatment. Because of the difficulty in collecting CT images of rectal tumors, we only trained and tested the proposed  segmentation network model in one rectal tumor database. In future, we will further train and test our proposed model in other medical CT image databases to further demonstrate it's performance.

					\end{document}